\documentclass[aps, pre, twocolumn, superscriptaddress, amsmath,amssymb,floatfix]{revtex4}

\newif\ifpdf\ifx\pdfoutput\undefined\pdffalse\else\pdfoutput=1\pdftrue\fi

\usepackage{graphicx}
\usepackage{bm}
\usepackage{epsfig}

\newcommand{\p}{^{0}}
\newcommand{\one}{^{1}}
\newcommand{\two}{^{2}}

\newcommand{\ronesig}{\rho\one(\sigma)}
\newcommand{\rtwosig}{\rho\two(\sigma)}

\begin{document}

\title{\bf Wetting transitions in polydisperse fluids}

\author{Matteo Buzzacchi}
\author{Nigel B. Wilding}
\affiliation{Department of Physics, University of Bath, Bath BA2 7AY, United Kingdom.}
\author{Peter Sollich}
\affiliation{King's College London, Department of Mathematics, Strand,
London WC2R 2LS, United Kingdom.}

\date{\today}

\begin{abstract} 

The properties of the coexisting bulk gas and liquid phases of a
polydisperse fluid depend not only on the prevailing temperature, but
also on the overall parent density. As a result, a polydisperse fluid
near a wall will exhibit density-driven wetting transitions inside the
coexistence region. We propose a likely topology for the wetting phase
diagram, which we test using Monte Carlo simulations of a model
polydisperse fluid at an attractive wall, tracing the wetting line
inside the cloud curve and identifying the relationship to prewetting.


\end{abstract} 
\maketitle
\setcounter{totalnumber}{10}

Complex fluids such as colloidal dispersions, polymers and liquid
crystals are inherently polydisperse, that is their constituent
particles exhibit continuous variation in terms of some attribute such
as their size, shape or charge. The presence of polydispersity in these
systems can profoundly influence their thermodynamic and processing
properties \cite{LARSON99,FREDRICKSON,SOLLICH02}. Recent work has shown
that the bulk phase behavior of polydisperse fluids can be surprisingly
rich, exhibiting novel features such as polydispersity-induced  phase
transitions in polymers \cite{GHOSH03}, reentrant melting in colloids
\cite{BARTLETT99}, and a sensitivity of coexistence properties to the
presence of rare large particles \cite{WILDING05}.  This richness is
traceable to {\em fractionation} effects \cite{SOLLICH02}: at
coexistence the distribution of the polydisperse attribute varies from
phase to phase. 

Notwithstanding the progress in elucidating the role of polydispersity
in determining bulk phase behaviour, little is known regarding
its effects on the {\em wetting} behavior of complex fluids. Besides
being a matter of great fundamental interest (see e.g.\
\cite{WIJTING03,AARTS04,FORSMAN05}) an understanding of this behaviour
is crucial for technological applications such as spin coating and
ink-jet printing of polymer or colloidal films for devices and
displays \cite{HERIOT05}.

This Letter broaches the subject of wetting in polydisperse fluids.
Starting from heuristic arguments based on the restricted case of a
binary mixture, we propose likely wetting scenarios for
polydisperse mixtures. These we test using detailed Monte Carlo
simulations of a model fluid at an attractive wall. Our results reveal a
novel form of wetting transition occurring {\em inside} the coexistence
region, which is driven by changes in {\em density} and/or temperature,
rather than by temperature alone as in the case of monodisperse
systems. We study the locus of wetting transitions in the phase diagram
and how it depends on the strength of the wall-fluid attraction. For
strongly attractive walls we further identify a prewetting transition
within the stable vapor phase region.

The state of a polydisperse system (or any of its phases) is
described by a density {\em distribution} $\rho(\sigma)$, with
$\rho(\sigma)d\sigma$ the number density of particles in the range of
$\sigma\ldots \sigma+d\sigma$; for e.g.\ size polydispersity, $\sigma$
would be the particle diameter. In the most commonly encountered
experimental situation, the form of the overall or ``parent''
distribution, $\rho\p(\sigma)$, is fixed by the synthesis of the fluid,
and only its scale can vary depending on the proportion of the sample
volume occupied by solvent. One can then write $\rho\p(\sigma)=n\p
f\p(\sigma)$ where $f\p(\sigma)$ is the normalized parent shape function
and $n\p=N/V$ the overall particle number density; as the latter is
varied, $\rho\p(\sigma)$ traces out a ``dilution line'' in density
distribution space. The phase diagram is then spanned by $n\p$ and the
temperature $T$.

Phase separation in a polydisperse system occurs within a region of
$n\p$ and $T$ delineated by the cloud curve, which records the
temperature $T$ where phase separation first occurs for given $n\p$. Quite
generally, one finds that at two phase coexistence the parent
$\rho\p(\sigma)$ will split into two ``daughter'' phase distributions
$\ronesig$ and $\rtwosig$ with different shapes---this is the phenomenon
of fractionation. The daughter distributions are related to the parent
via a simple volumetric average:
$(1-\xi)\ronesig+\xi\rtwosig=\rho\p(\sigma)\:,$ where $1-\xi$ and $\xi$
are the fractional volumes of the two phases. In contrast to a
monodisperse system where the densities of the coexisting phases are
specified by the binodal and depend solely on temperature, full
specification of the coexistence properties of a polydisperse system
requires not only knowledge of the cloud curve, but also the dependence
of $\xi$, $\ronesig$ and $\rtwosig$ on $n\p$ and $T$ \cite{BUZZACCHI06}.

\begin{figure}[h]
\includegraphics[type=eps,ext=.eps,read=.eps,width=8.0cm,clip=true]{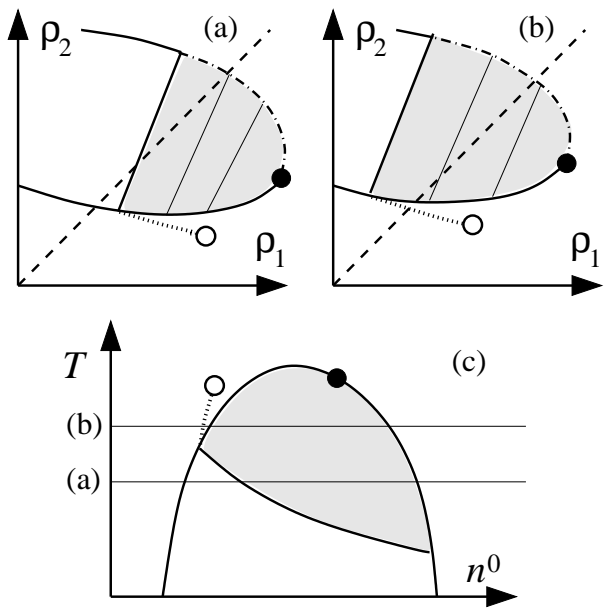} 

\caption{Schematic phase diagram of the binary mixture described in the
text. Thick solid line: coexistence boundary; thin solid lines:
tie-lines; grey areas: liquid (dash-dotted line) wets wall; dashed line:
dilution line; dotted line: prewetting line, with CP (open circle). Full
circle: bulk CP. (a) and (b): isothermal cuts at two different $T$ as
indicated in the full phase diagram (c) for fixed $\rho\p_1/\rho\p_2$.}
\label{fig:binary}
\end{figure}

To understand the effects of polydispersity on wetting transitions, we
turn first to the simplest case of a binary mixture with densities
$\rho_1$ and $\rho_2$ of two species of particles of different sizes;
these are the analog of $\rho(\sigma)$. Fig.~\ref{fig:binary}(a)
sketches an exemplary~\cite{sketch} isothermal cut through a bulk
phase diagram, showing a region of liquid-vapor phase separation with tie-lines that shrink to
zero at a critical point (CP). We assume that the particle-wall
interaction strength (averaged over the two species) is sufficient that
there is a finite region of coexistence, bordering the CP, for which the
liquid wets the wall (the wall is always wet {\em at} criticality
because of the rapid decrease of the liquid-vapor surface tension~\cite{DIETRICH});
a prewetting line extending into the one-phase region is also indicated.

The dilution line constraint of a fixed shape for $\rho\p(\sigma)$
reduces in the bidisperse case to a fixed ratio $\rho\p_1/\rho\p_2$. As
$n\p=\rho\p_1+\rho\p_2$ is increased from zero, the system follows the
dilution line into the coexistence region. For a given point on the
dilution line, the parent splits into two daughter phases located at the
ends of the tie-line which intersects this point; owing to
fractionation, the daughters lie {\em off} the dilution line. In the
example of Fig.~\ref{fig:binary}(a), one sees that as the system enters
the coexistence region the liquid does not initially wet the wall, but
as $n\p$ increases further, the properties of the daughter phases
change, and it starts to do so; crucially, this change occurs {\em
within} the coexistence region.  Fig.~\ref{fig:binary}(b) shows a
possible scenario at a higher temperature; here the phase coexistence
region has shrunk and thus the dilution line now crosses the prewetting
line in the one-phase region, indicating that a prewetting transition
will occur on increasing $n\p$. In  this case, the liquid daughters
always wet the wall, and so no further transitions occur inside the
coexistence region. Generalizing from these two fixed temperatures to a
full $n\p-T$ phase diagram, we infer the topology shown in
Fig.~\ref{fig:binary}(c); the wetting transitions inside the coexistence
region will occur for a range of temperatures and so define a wetting
line which we expect to link the low and high density sides of the
coexistence region. Provided this wetting line is first order where it
intersects the gas branch, it will connect to a prewetting line, as
shown.

We have not, of course, explored all possible phase diagram structures
here; even for bidisperse systems at constant $T$, other possibilities
could be envisaged. We are also not aware of any rigorous argument that
indicates that the wetting line must have a negative slope as in our
illustration. However, at least for weak wall interactions, where the
wetting line is expected to enclose only a small region around the CP,
its slope should follow that of the cloud curve there; in all the
examples of which we are aware, this slope is negative, and so we may
anticipate that the scenario we have outlined is quite general. 

To test the above ideas in a genuinely polydisperse setting, we have
deployed Monte Carlo simulation to study a model fluid in which the
polydisperse attribute $\sigma$ is the particle diameter. Particles
interact via a Lennard-Jones potential of the form:
\begin{equation}
u_{ij}=\epsilon_{ij}\left[\left({\sigma_{ij}}/{r_{ij}}\right)^{12}
-\left({\sigma_{ij}}/{r_{ij}}\right)^{6}\right]
\label{uij}
\end{equation}
with $\epsilon_{ij}=\sigma_i\sigma_j\epsilon/\bar\sigma^2$, where
$\epsilon$ and $\bar\sigma$ set our energy and length scale units
respectively, while
$\sigma_{ij}=(\sigma_i+\sigma_j)/2$ and $r_{ij}=|{\bf r}_i-{\bf
r}_j|$. The potential was truncated for $r_{ij}>2.5\sigma_{ij}$ and no
tail corrections were applied. The diameters
$\sigma$ are drawn from a (parental) Schulz distribution
$f\p(\sigma)\propto\sigma^z\exp\left[(z+1)\sigma/\bar{\sigma}\right]$,
with mean diameter $\bar{\sigma}$. We
chose $z=50$, corresponding to a moderate degree of polydispersity: the
standard deviation of particle sizes is $\delta\equiv
1/\sqrt{z+1}\approx 14\%$ of the mean. The distribution $f\p(\sigma)$
was limited to within the range $0.5\bar\sigma<\sigma<1.4\bar\sigma$. 

The bulk liquid-vapor phase behaviour of the model of eq.~(\ref{uij})
has previously been obtained~\cite{WILDING05} within the grand canonical
ensemble (GCE) using bespoke computational techniques designed to ensure
that the ensemble averaged density distribution always equals a desired
parent form, by controlling an imposed chemical potential distribution
$\mu(\sigma)$. Also implemented was a recently proposed strategy
\cite{BUZZACCHI06} which ensures that the two phases appear with equal
probabilities in the GCE simulations. This results in finite-size
corrections to the limiting bulk properties which are exponentially
small in the system size -- a feature which, it transpires, is crucial
for accurately obtaining the wetting properties corresponding to a
semi-infinite system. The study of ref.~\cite{WILDING05} yielded ({\em
inter alia}) accurate estimates for the cloud curve and the form of
$\mu(\sigma)$ at all points within the coexistence region. The CP was
found to be located not at the temperature maximum of the cloud curve,
but at higher density and lower temperature, as is also observed in
experimental studies of polydisperse systems \cite{KITA97}.

Equipped with knowledge of the bulk phase behavior of our model, we turn
now to address its wetting properties \cite{DIETRICH} by introducing a
single attractive wall to our system. Our simulation box has dimensions
$L_x\times L_y\times L_z$, with $L_x=L_y=15\bar\sigma$ and
$L_z=30\bar\sigma$ or $L_z=40\bar\sigma$, and is periodic in the $x$ and
$y$ directions. A fluid particle interacts with the attractive wall at
$z=0$ via a potential 
\begin{equation}
 U_{iw}=\epsilon_{iw}\left[\left({\sigma_{iw}}/{z_{iw}}\right)^{10}
-\left({\sigma_{iw}}/{z_{i}}\right)^{4}\right]\:,
\end{equation}
with $\epsilon_{iw}=\epsilon_w\bar\sigma\sigma_i/{\bar\sigma}^2$ and $\sigma_{iw}=(\sigma_i+\bar\sigma)/2$.  The wall at $z=L_z$ is
completely hard.

In a canonical setting, the wall is ``wet'' at a given coexistence state
point if the free energy is minimized when liquid is in contact with the
wall, rather than gas. In our GCE simulations, we locate wetting points
for a given wall strength $\epsilon_w$ by imposing the $\mu(\sigma)$
appropriate to some prescribed bulk values of $n\p$ and $T$; this
corresponds to the conditions of a semi-infinite geometry. We then
record the ensemble averaged density profile $\bar{n}(z)=\int d\sigma
\bar{\rho}(z,\sigma)$ and the probability distribution $p(n)$ of the
fluctuating total density $n=\int d\sigma\,dz\,\rho(z,\sigma)$;
both of these are accumulated as histograms, one over $z$ and the other
over $n$. We repeat these measurements, for a range of bulk state
points in which $n\p$ is increased isothermally from the
stable vapor region to values within the bulk coexistence region.
Formally, wetting occurs within the GCE when the adsorption
$\Gamma(n\p)=\int_0^\infty (\bar{n}(z)-n_v)dz$ diverges, with $n_v$ the
vapor density. For systems of finite size, the wetting transition is
identifiable as the value of $n\p$ at which the instantaneous profile
$n(z)$ first becomes freely fluctuating. This leads to a linear decay of
the $\bar{n}(z)$ along the $z$-direction; it is also signalled
by the disappearance of a ``vapor peak'' in the distribution of the
overall density $p(n)$ \cite{MACDOWELL06}. 

As an independent consistency check on this approach for determining
wetting points, we have, for a broad selection of our state points, also
implemented the distinct method described in ref.~\cite{MUELLER98}. Here
direct measurements (in geometries different to that considered above)
of liquid-vapor, vapor-wall and liquid-wall surface tensions, together
with Young's equation, $\gamma_{LV}\cos\theta=\gamma_{VW}-\gamma_{LW}$,
serve to estimate the contact angle $\theta$. Wetting points follow as
those combinations of $n\p$ and $T$ where $\theta$ vanishes. The
requisite surface tensions were extracted from
the form of $p(n)$ (via multicanonical simulations~\cite{BERG}) for both
a bulk (periodic) system and a system with two identical attractive
walls (at $z=0$ and $z=L_z$) as described in ref.~\cite{MUELLER98}. In
all cases, good agreement was found with the estimates of wetting points
obtained from the observations of $\bar{n}(z)$ and $p(n)$, described
above, indicating that the effect of the hard wall at $z=L_z$ on the
freely fluctuating form of $\bar{n}(z)$ does not engender significant
finite-size corrections to estimates of wetting points.

\begin{figure}[h]
\includegraphics[type=eps,ext=.eps,read=.eps,width=8.5cm,clip=true]{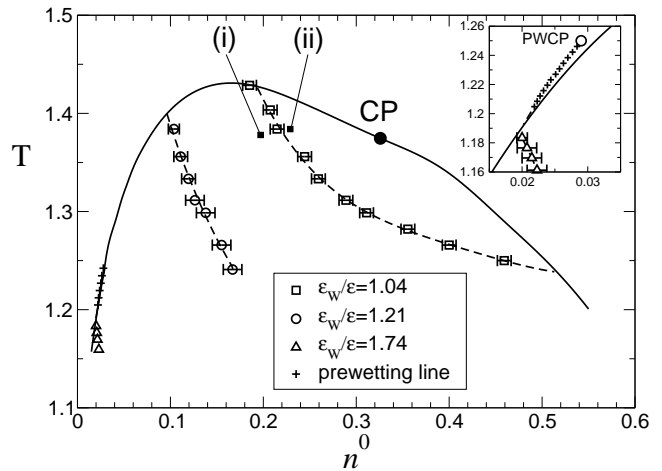}
\caption{Wetting phase diagram for three wall strengths. Full line:
cloud curve obtained in Ref.~\cite{WILDING05}. Symbols: measured points on the wetting
line; dashed lines guide the eye; small solid squares: state points
referred to in Fig.~\ref{fig:profiles}. Inset: prewetting line
observed for $\epsilon_w/\epsilon=1.74$.}
\label{fig:wpd}
\end{figure}

\begin{figure}[h]
\includegraphics[type=eps,ext=.eps,read=.eps,width=8.5cm,clip=true]{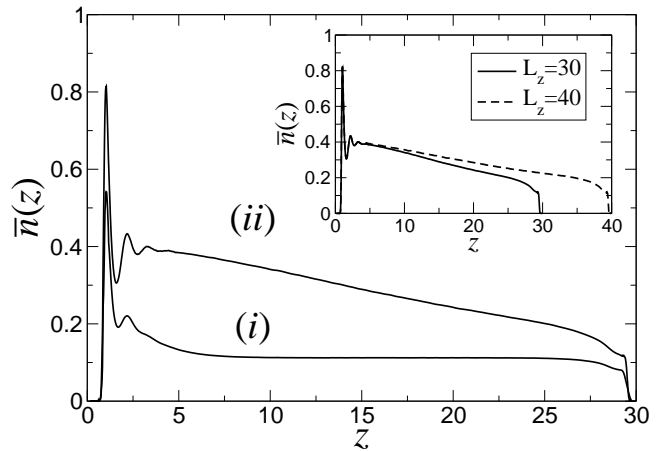}
\caption{Profiles $\bar{n}(z)$ for the state points indicated in
Fig.~\protect\ref{fig:wpd}, see text. The inset shows
the effect of increasing $L_z$ on the profile of the freely fluctuating film.}
\label{fig:profiles}
\end{figure}

Fig.~\ref{fig:wpd} shows our measurements of the locus of wetting
transitions in the $n\p-T$ plane, together with the cloud curve, for
three choices of the wall strength $\epsilon_w$.  Considering first the
weakest wall, $\epsilon_w=1.04$, we find that the line of
wetting points traverses the coexistence region, intersecting the cloud
curve discontinuously at its high and low density branches. The line has
a negative gradient and encloses a portion of the coexistence region
containing the CP. For coexistence points inside this region
the wall is wet, while for points outside, it is dry.
Fig.~\ref{fig:profiles} shows density profiles $\bar{n}(z)$ for two
state points slightly on either side of the wetting line. When the wall
is dry, the profile is gas like, except for a small enhancement at
the wall. However, when the wall is wet, the liquid-gas interface
fluctuates freely and the ensemble averaged profile is sloped because
the maximum extent of the fluctuations is limited by the hard wall at
$z=L_z$. For a system infinite in the $z$ direction, the time averaged
profile of the wet wall would become flat as a function of $z$. We have
verified that the profile indeed becomes flatter as we increase $L_z$
(inset Fig.~\ref{fig:profiles}), although no accompanying systematic
changes in the locus of the wetting line were observed.

\begin{figure}[h]
\includegraphics[type=eps,ext=.eps,read=.eps,width=8.2cm,clip=true]{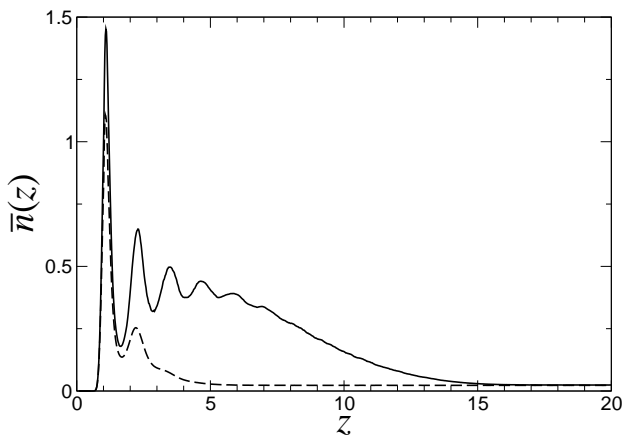}
\caption{Typical profiles $\bar{n}(z)$ for showing coexisting thin and thick films
at the prewetting transition for $L_z=30\bar\sigma$, see text.}
\label{fig:prewet}
\end{figure}

For the intermediate wall strength $\epsilon_w=1.21$, the region of
coexistence for which the wall is wet appears to be larger than for the
weak wall, although we were unable to follow the wetting line right across the
coexistence region due to the computational burden of simulating at very
low $T$ and large $n\p$. For both the weak and intermediate wall
strengths, no sign of prewetting was discernible in the stable vapor phase
region. However a clear signature of prewetting was found for the
strongest wall strength, $\epsilon_w=1.74$, at low $T$ in the stable
vapor phase region near where the wetting line (of which we were able to
determine only a small segment) intersects the gas branch of the cloud
curve. The inset of Fig.~\ref{fig:wpd} depicts the measured line of
prewetting transitions, obtained from the points at which the film
thickness jumps from a thin to thick finite value, as shown in
Fig.~\ref{fig:prewet}. We note that the prewetting line appears to
intersect the cloud curve tangentially, as is mandated thermodynamically
\cite{HAUGE}.

Although the prewetting line represents a true (surface) phase
transition, we note that no fractionation effects (such as splitting of
the prewetting line into separate cloud and ``shadow'' curves -- see
Ref.~\cite{SOLLICH02}) are expected because neither of the coexisting
phases are `macroscopic' in the sense that their volume scales like the
system volume. We further note that the observation of prewetting is
evidence of a first order wetting transition \cite{DIETRICH}, while its
absence at smaller wall strengths could indicate that critical wetting
occurs here. Intriguingly, there seem no grounds to rule out, {\em a
priori}, a scenario in which the wetting transition changes from first
order to continuous wetting at some point {\em along} the wetting line,
the two regimes being separated by a tricritical wetting point.


The new feature of the wetting phase diagram of Fig.~\ref{fig:wpd},
compared to its monodisperse counterpart, is the occurrence of wetting
transitions at constant $T$ driven by changes in parent density.
The wetting line delineates a subregion of the coexistence region for
which the wall is wet, and which contains the critical point. We expect
this subregion to shrink {\em onto} the critical point as the wall
strength is progressively weakened. We note that one can view the
wetting line in the $n\p-T$ plane as the analog of the
wetting temperature isotherm $T=T_w$ (a horizontal line) in the $n-T$
phase diagram of a monodisperse system. The variation of daughter
distributions with parent density in a polydisperse system causes this
line to `tilt'. The simulation results confirm our heuristic arguments
to this effect based on a simple bidisperse example.

Finally, with regard to the wider impact of our findings, we expect
these to be relevant to the surface phase behavior of all complex fluids
that exhibit significant polydispersity. More specifically, the
possibility of density driven wetting transitions could aid an
understanding of the factors controlling the morphology of spin-coated
or ink-jet printed thin films, which arises from the interplay of
wetting and phase separation \cite{HERIOT05}. During these production
processes, the solvent is progressively removed, thereby increasing the
solute density, and possibly leading to the type of wetting transitions
we have described.








\begin{thebibliography}{99}

\bibitem{LARSON99} R.~G. Larson, {\em The Structure and Rheology of
Complex fluids} (Oxford University Press, New York, 1999).

\bibitem{FREDRICKSON} G.~H. Fredrickson, Nature (London) {\bf 395}, 323
(1998).

\bibitem{SOLLICH02} For a review, see P. Sollich, J. Phys. Condens. Matter {\bf 14}, R79 (2002).

\bibitem{GHOSH03} K. Ghosh and M. Muthukumar, Phys. Rev. Lett. {\bf 91},
158303 (2003).

\bibitem{BARTLETT99} P. Bartlett and P.~B. Warren, Phys. Rev. Lett. {\bf 82}, 1979 (1999).

\bibitem{WILDING05} N.~B. Wilding {\em et al}, Phys. Rev. Lett. {\bf 95}, 155701
(2005); N.~B. Wilding {\em et al}, J. Chem. Phys. {\bf 125}, 014908 (2006).

\bibitem{WIJTING03} W.~K. Wijting, N.~A.~M. Besseling and M.~A. Cohen
Stuart, Phys. Rev. Lett. {\bf 90}, 196101 (2003).

\bibitem{AARTS04} D.~G.~A.~L. Aarts {\em et al}, J. Chem. Phys. {\bf 120},
1973 (2004).

\bibitem{FORSMAN05} J. Forsman and C.~E. Woodward, Phys. Rev. Lett. {\bf 94}, 118301 (2005)

\bibitem{HERIOT05} S.~Y. Heriot and R.~A.~L. Jones, Nature Materials, {\bf 4},
782 (2005); B.-J. de Gans and U.S. Schubert, Langmuir {\bf 20}, 7789 (2004).

\bibitem{BUZZACCHI06} M. Buzzacchi {\em et al}, Phys. Rev. E{\bf 73}, 046110 (2006).


\bibitem{sketch} We assume that species 2 (taken as the larger
  particles) has stronger attractive interactions and so phase separates on
  its own (vertical axis) at the chosen $T$ while the pure species 1
  fluid (horizontal axis) does not. With interaction strengths chosen
  in this way, species 2 particles will typically accumulate in the
  liquid, with its shorter interparticle distances. The tielines in
  the sketch therefore cross any dilution line from below, moving from
  smaller values of $\rho_2/\rho_1$ at the gas end to larger ones for
  the coexisting liquid.

\bibitem{DIETRICH} For a review of wetting, see eg. S. Dietrich, {\em
Phase Transitions and Critical Phenomena}, edited by C. Domb and J.L.
Lebowitz (Academic, London, 1988), Vol 12.

\bibitem{KITA97} See eg. R. Kita {\em et al}, Phys. Rev. E {\bf 55}, 3159
(1997).

\bibitem{MACDOWELL06} L.~G. MacDowell and M. M\"{u}ller, J. Chem. Phys.
{\bf 124}, 084907 (2006).

\bibitem{MUELLER98} M. M\"{u}ller and K. Binder, Macromolecules {\bf 31}, 8323 (1998).

\bibitem{BERG} B.A. Berg and T. Neuhaus, Phys. Rev. Lett. {\bf 68}, 9 (1992).

\bibitem{HAUGE} E.H. Hauge and M. Schick, Phys. Rev. B{\bf 27}, 4288
(1983).




\end{thebibliography}
\end{document}